\documentclass[twocolumn]{jpsj2}

\def\runtitle{Quantum phase transitions of the $S=1$ Shastry-Sutherland
model}
\def\runauthor{Akihisa \textsc{Koga}, Norio \textsc{Kawakami} and 
Manfred \textsc{Sigrist}}

\title{
Quantum phase transitions of the $S=1$ Shastry-Sutherland model
}

\author{%
Akihisa Koga, Norio Kawakami and
Manfred Sigrist$^{1}$
}

\inst{
Department of Applied Physics, Osaka University, Suita, 
Osaka 565-0871, Japan\\
$^1$Theoretische Physik, ETH-H\"onggerberg, Z\"urich 8093, Switzerland 
}

\recdate{\today}

\abst{%
We investigate quantum phase transitions of the two-dimensional $S=1$ 
Shastry-Sutherland model, which is characterized by the frustrated 
orthogonal-dimer structure, by means of the exact diagonalization method
and the Ising-type series expansion method. 
We clarify how distinct spin gap phases realized in the chain system
are adiabatically connected to those in the two-dimensional
Shastry-Sutherland model. 
The effect of single-ion anisotropy is also discussed.
}

\kword{%
Shastry-Sutherland model, Valence Bond Solid, exact diagonalization,
Ising expansion
}

\begin{document}
\sloppy
\maketitle

\section{Introduction}
\begin{figure}[htb]
\begin{center}
\includegraphics[width=7cm]{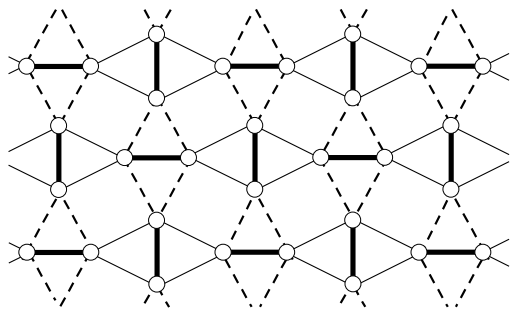}
\end{center}
\caption{Orthogonal-dimer structure: the bold, thin and broken lines
 represent the exchange couplings $J, J'$ and $J''$.  Note that the dimers
indicated by the bold lines are orthogonal to each other. When $J''=0$, the 
system is reduced to the orthogonal-dimer chain.
}
\label{fig:model}
\end{figure}

Frustrated spin systems have attracted much theoretical
attention over many years. Interesting experimental realizations have been
recently found in the transition-metal oxides
$\rm SrCu_2(BO_3)_2$\cite{Kageyama,SciKodama} and $\rm
Nd_2BaZnO_5$,\cite{Kageyama3} 
where the magnetic ions 
$\rm Cu^{2+}$ and $\rm Nd^{3+}$ sit on the orthogonal-dimer structure
(see Fig. \ref{fig:model}).  
In the compound $\rm SrCu_2(BO_3)_2$, a dimer singlet 
phase is realized
due to the strong antiferromagnetic couplings for 
dimer bonds.  A number of extraordinary magnetic properties
were observed such as magnetization plateaus, dispersionless
excited states,  which have 
stimulated further experimental
\cite{Onizuka,Kodama,Nojiri,Kageyama2,Cepas,Wolf} 
and theoretical
\cite{Shastry,Miyahara,Koga,Kogachain,Takushima,Sigrist,Momoi,Fukumoto,
Knetter,Misguich,Chung,Zheng,Koga3D,KogachainH,Ivanov,Totsuka}
studies.
On the other hand, in the compound $\rm Nd_2BaZnO_5$ \cite{Kageyama3} 
the neodium ion has a large magnetic moment
$J=9/2$, leading to an antiferromagnetically ordered state 
below $T_N=2.4K$.

Magnetic properties of the above compounds may be described by the
Heisenberg model on the square lattice with some diagonal bonds, 
which is referred to as the Shastry-Sutherland model.\cite{Shastry}
This class of quantum spin models have the striking property
that the direct product of independent dimers gives an exact 
eigenstate of the system, which can be the ground state
for a certain range of parameters.
Various aspects have been discussed  for the $S=1/2$ spin system,
such as quantum phase transitions,
\cite{Miyahara,Koga,Kogachain,Takushima,Sigrist,Knetter,Chung,Zheng,
Koga3D,KogachainH,Ivanov}  the
correlated hopping of triplet excitations, 
\cite{Momoi,Totsuka,Fukumoto,Knetter} the 
 plateau formation in the magnetization curve,
\cite{Miyahara,Momoi,Fukumoto,Misguich,KogachainH,Ivanov} and so on. 
However, magnetic properties of higher spin models ($S>1/2$) have not
been addressed so far, apart from a simple-minded approach by
Shastry and Sutherland in their pioneering work.\cite{Shastry}
Therefore, it is desirable to discuss how the
competing exchange interactions affect the ground
state properties of a higher-spin $(S>1/2)$ orthogonal-dimer model.

In a previous paper,\cite{Kogachain} 
we have investigated the one-dimensional version of the
orthogonal-dimer spin model with an arbitrary spin, and have shown that
first-order quantum phase transitions occur $(2S)$ times
when the ratio of two competing antiferromagnetic exchange interactions
are changed. In particular,
in the $S=1$ system, a non-trivial spin gap phase 
exists between the dimer and the plaquette phases. 
In the present paper, we investigate the ground-state phase diagram of 
the two-dimensional (2D) $S=1$ Shastry-Sutherland model, and 
clarify  how the above spin-gap phases in 1D are 
adiabatically connected to those in the 2D model
by introducing interchain couplings.

The paper is organized as follows. In Sec. \ref{sec:Model},
we introduce the $S=1$ Shastry-Sutherland model, and discuss how the
quantum phase transitions are affected by spatially
 anisotropic exchange 
couplings. 
In Sec. \ref{sec:anisotropy}, we also discuss
the effect of single-ion anisotropy, which is important
for some materials. 
A brief summary is given in Sec. \ref{sec:summary}.

\section{Phase diagram of the isotropic model}\label{sec:Model}

We consider the generalized Shastry-Sutherland model,
\begin{eqnarray}
H=\sum_{(i,j)}J_{ij}\mib{S}_i\cdot\mib{S}_j+D\sum_{i}
\left(S_i^z\right)^2,\label{eq:Hami} 
\end{eqnarray}
where $\mib{S}_i$ denotes the $S=1$ spin operator on the $i$th site,
and
 $J_{ij}=J, J'$ and $J''$ represent the intra-dimer, the inter-dimer
and the inter-chain couplings, which are all assumed to be
antiferromagnetic. The single-ion anisotropy is denoted as $D(<0)$.
A schematic view of the model is given in Fig. \ref{fig:model}. 

In order to see
 what kind of spin-gap phases or the magnetically ordered phases
are realized in the 2D Shastry-Sutherland model $(J'=J'')$, 
we perform the exact diagonalization ($N=4\times4$ system) of the 
above generalized spin model. We focus on the 
isotropic model with $D=0$ in this section.
\begin{figure}[htb]
\begin{center}
\includegraphics[width=7cm]{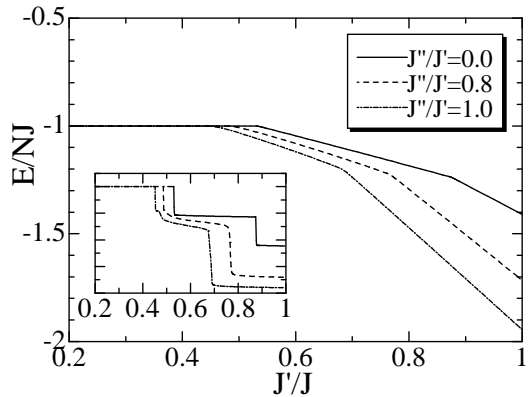}
\end{center}
\caption{Ground state energy of the isotropic model 
as a function of $J'/J$ obtained by the
 exact diagonalization of the small system 
 with $4\times4$ sites. From up to down,
 $J''/J'=0.0, 0.8$ and $1.0$. Inset shows the derivative of the ground
 state energy, from which we can clearly see the nature of the 
 first-order phase transition.
}
\label{fig:ene}
\end{figure}

\begin{figure}[htb]
\begin{center}
\includegraphics[width=7cm]{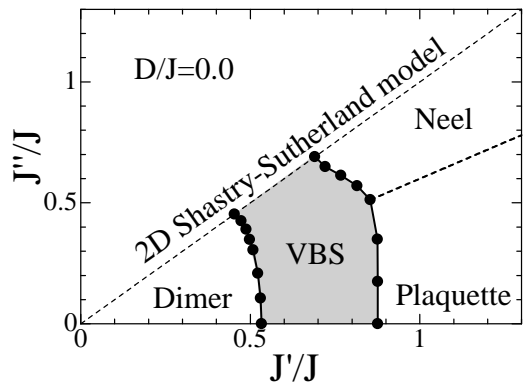}
\end{center}
\caption{
Phase diagram for the isotropic model, $D=0$.
The bold lines  with solid circles represent the phase
boundary at which  the first-order phase transition occurs. 
The broken line between the plaquette phase and the 
antiferromagnetic (AF) phase indicates the approximate location of the
phase boundary.
}
\label{fig:phase}
\end{figure}

The ground state energy computed  for the isotropic model
($D=0$) is shown in Fig. \ref{fig:ene}.
It is seen that two unambiguous cusps appear in the energy diagram,
irrespective of the choice of the exchange couplings, implying 
 that the first-order quantum phase transition occurs twice when
 $J'/J$ increases.
The reason why we obtain such clear cusp structures even for the 
small $4\times 4 $ system is closely related to the 
characteristic orthogonal-dimer 
structure, for which the direct product of  spatially decoupled dimers  
gives an exact eigenstate, making the spin correlation length 
 extremely short.
The phase diagram determined from the cusp structure of the ground-state
energy is shown in Fig. \ref{fig:phase}.

\begin{figure}[htb]
\begin{center}
\includegraphics[width=8cm]{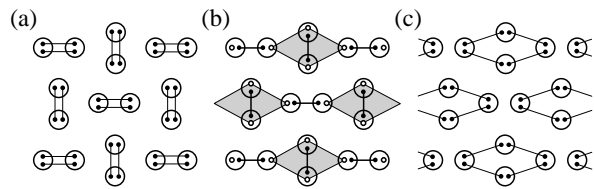}
\end{center}
\caption{VBS picture of the dimer phase (a), 
the frustration-induced phase (b) and the plaquette phase (c)
for the generalized $S=1$ Shastry-Sutherland model. In this 
description, it is assumed that the 
original $S=1$ spin is decomposed into two 
$S=1/2$ spins denoted by dots, and each bond connects 
the decoupled $S=1/2$ spins
to make singlet. Note that the system is covered by the 
periodic array of dimers and plaquettes
for the intermediate VBS phase (b).}
\label{fig:VBS}
\end{figure}

Let us start with the case of $J''=0$, which 
is equivalent to the orthogonal-dimer spin
chain studied previously.\cite{Kogachain}  In the chain model,
 the dimer- (plaquette-) 
singlet phase is realized for  $J'/J<0.56$ ($J'/J>0.88$), 
both of which are characterized by the 
disordered ground state with a triplet excitation gap.
In the intermediate region ($0.55<J'/J<0.88$), 
another singlet phase with a triplet excitation gap is stabilized by
strong frustration,  which is known to be
 topologically equivalent to the  Haldane-gap
 phase.\cite{Kogachain}  As is discussed
 in detail below, 
this frustration-induced state is composed of periodic 
alignment of dimer- and plaquette-singlets of 
$S=1/2$ decoupled spins, 
 so that it is
regarded as a kind of Valence Bond Solid (VBS) state
(see Fig. \ref{fig:VBS}).

 We naively expect that these spin-gap phases may be 
 unstable in the presence of the interchain coupling since 
 such quasi-one dimensional $S=1$ spin system  is usually 
driven to the antiferromagnetically ordered phase.
\cite{Sakai,KogaHaldane,KogaD,Kim,Matsumoto,Kawaguchi} 
However, according to 
the present exact diagonalization study, we find 
that two of the spin gap 
 phases, i.e. the dimer phase  and the intermediate VBS phase, 
 are stable against the interchain couplings, and persist even in the 
Shastry-Sutherland model ($J''=J'$).
In particular, it is remarkable that the nontrivial VBS phase 
exists even in the 2D Shastry-Sutherland model.

 The above three
phases are clearly distinguished from each other according to the
topological nature specified by the VBS description,\cite{VBS} 
where each  singlet ground  state is
represented by the assembly of the singlet bonds between the decomposed
$S=1/2$ spins. 
 Shown in Fig. \ref{fig:VBS} is the VBS description of these spin 
 gap phases.
In the dimer phase, two singlet bonds are located on the strong exchange
coupling $J$. Then the assembly of the singlet dimers 
gives  the exact ground
state of the Hamiltonian eq. (\ref{eq:Hami}).
 On the other hand, in the frustration-induced VBS 
  phase,  one of the decomposed
spins at each site is connected to the nearest neighbor spin to form the
singlet-dimer. Another decomposed spin is connected to other three spins
to form the plaquette singlet [Fig. \ref{fig:VBS}(b)]. This  phase may 
be referred to as the
"dimer-plaquette VBS phase", since it is composed of periodically 
alternating  dimer and  plaquette singlets.

Somewhat subtle is the stability of the plaquette phase against 
the antiferromagnetic ordered phase.  Since the plaquette
state forms a four-spin singlet network [Fig.\ref{fig:VBS}(c)], 
whose wavefunction is rather extended spatially,
it will show a second-order phase transition to the ordered  phase
 as  the interchain coupling
is increased. This is indeed the case for the $S=1/2$ model, for which
the plaquette phase becomes unstable at a certain  
interchain coupling, and undergoes a second-order 
phase transition to the magnetically ordered phase 
except for a very narrow window of the choice of
 the exchange couplings.\cite{Koga,Takushima,Sigrist} 
In the present $S=1$ case, the plaquette phase is even 
more unstable compared with the $S=1/2$ case
since the $S=1$ system favors the
 ordered state. Therefore we believe that
 the plaquette phase may 
completely disappear in the Shastry-Sutherland model ($J''=J'$),
although it is difficult to determine the phase boundary
of the second-order transition
numerically from the small system.
The phase boundary between the plaquette phase and 
the antiferromagnetic phase is shown as a guide to eyes
in Fig. \ref{fig:phase}.

\section{Effects of single-ion anisotropy}\label{sec:anisotropy}

We now discuss the effect of single-ion anisotropy, which sometimes
plays an important role in stabilizing the magnetically ordered
 state in real materials with higher spins.
It is naively expected that such anisotropy has a tendency to 
drive the system to the magnetically ordered phase discussed in 
the previous section.  However,  it 
should be noticed that the magnetic phase 
stabilized by single-ion anisotropy may be
distinct from the magnetic phase in the previous section
due to the competing interactions.

\begin{figure}[htb]
\begin{center}
\includegraphics[width=7cm]{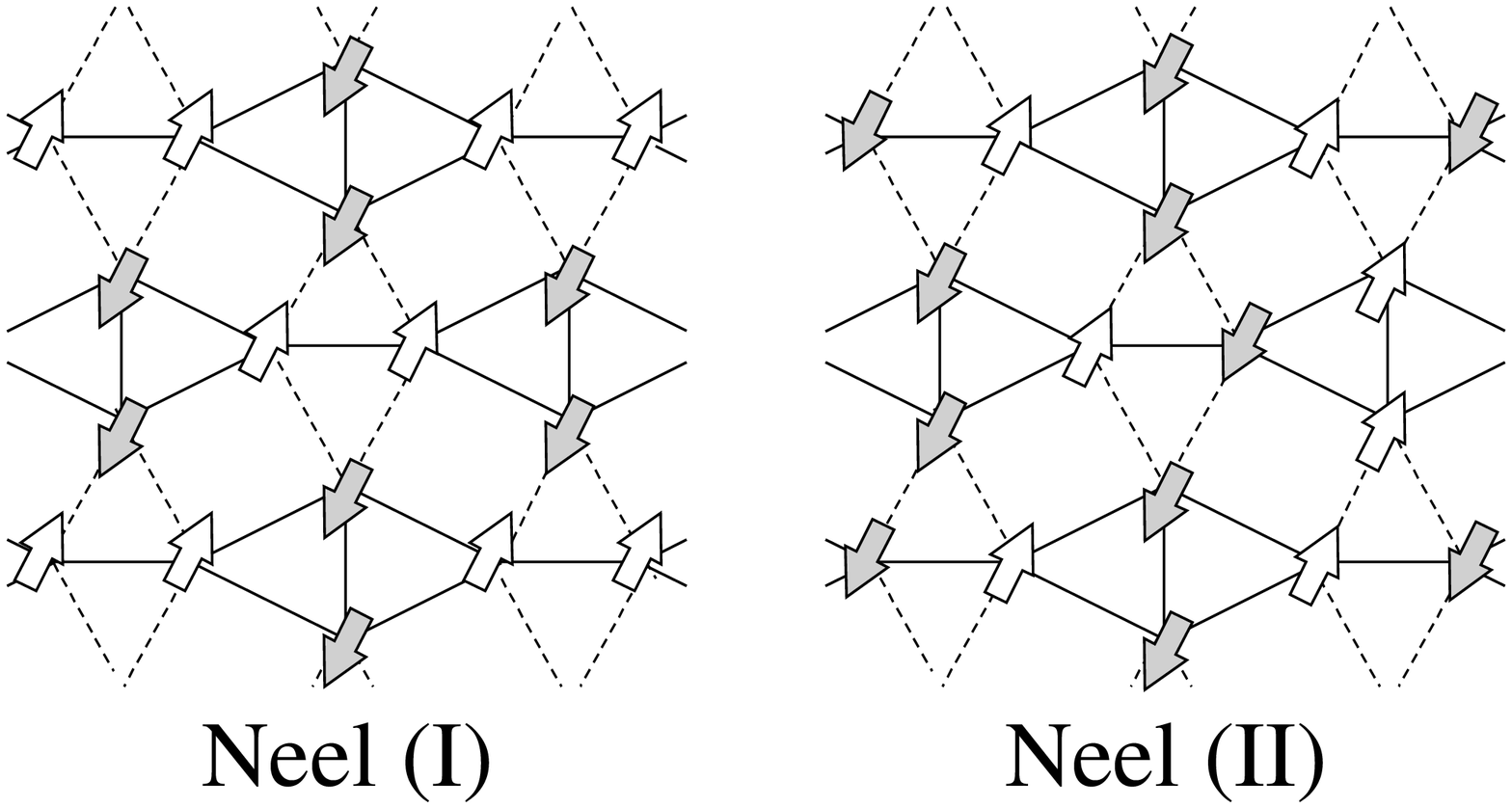}
\end{center}
\caption{Spin configurations for 
the Neel ordered phases (I) and (II).
}
\label{fig:neel}
\end{figure}

We have verified that there are indeed two kind of 
ordered phases in our model, which are labeled by Neel (I) and 
 Neel (II), whose spin configuration 
is schematically drawn in Fig. \ref{fig:neel}.
The phase (I) is the Neel ordered phase introduced above in this
paper, characterized by order moments staggered among the dimers. 
In contrast the phase (II) corresponds to a staggering along the
chains and is expected for 
 the Neel phase stabilized  by single-ion anisotropy. Obviously, the
 gradual increase of the interchain coupling $ J''/J $ yields an
 incompatible situation for the staggered ordering along the chain and 
the phase (I) becomes favorable.

\begin{figure}[htb]
\begin{center}
\includegraphics[width=7cm]{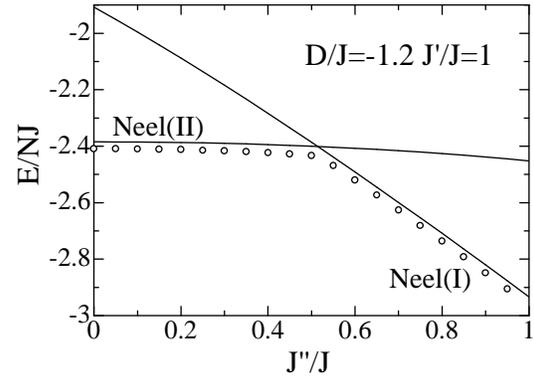}
\end{center}
\caption{Ground state energy as a function of $J''/J$ for the 
anisotropic model. Solid lines are
 obtained by the Ising expansion up to the third order in $\lambda$. The
 open circles are obtained by the exact diagonalization calculation
 for the system of $N=4\times4$.
}
\label{fig:ene-D}
\end{figure}

The phase transition between these two 
distinct ordered phases should be of first-order, since there is no
symmetry-hierarchy between them. 
To clarify this point, we make use of the Ising expansion,\cite{series}
starting from two different spin  configurations shown 
in Fig. \ref{fig:neel}. The accuracy of this 
approach will be further confirmed below by comparing
the results with those of the exact diagonalization calculation.
To perform the series expansion, we now divide the original 
Hamiltonian into two parts as
\begin{eqnarray}
H&=&H_0+\lambda H_1,\\
H_0&=&\sum J_{ij}S_i^zS_j^z+D\sum\left(S_i^z\right)^2,\nonumber\\
H_1&=&\sum J_{ij}\left(S_i^xS_j^x+S_i^yS_j^y\right).
\end{eqnarray}
Then the ground state energy of each ordered state is expanded in
$\lambda$ as,
\begin{eqnarray}
\frac{E_I}{N}&=&\frac{J}{2}-J'-J''+D-\left(
\frac{{J'}^2}{E_{I,1}}+
\frac{{J''}^2}{ E_{I,2}}\right)\lambda^2\nonumber\\
&+&J\left[\left(\frac{J'}{ E_{I,1}}\right)^2+
\left(\frac{J''}{ E_{II,2}}\right)^2\right]\lambda^3
+O\left(\lambda^4\right)\\
\frac{E_{II}}{N}&=&-J'+D-\left(\frac{{J'}^2}{ E_{II,1}}+
\frac{{J''}^2}{2 E_{II,2}}+\frac{J^2}{4 E_{II,3}}\right)\lambda^2\nonumber\\
&+&J\left[\left(\frac{J'}{E_{II,1}}\right)^2+
\frac{{J''}^2}{E_{II,2}E_{II,3}}\right]\lambda^3+O\left(\lambda^4\right),
\end{eqnarray}
where $N$ is a total number of sites, 
$E_{I,1}=-2J+3J'+4J''-2D, E_{I,2}=-2J+4J'+3J''-2D, 
E_{II,1}=3J'-2D, E_{II,2}=4J'-J''-2D$ and $E_{II,3}=J+4J'-2D$.
It is found that the series coefficients in the expansion
approach zero in both cases  when $D\rightarrow-\infty$, from which 
we can see that the Ising expansion is an 
appropriate method to discuss 
the magnetic properties of the model quantitatively.
We show the energy computed for each phase $(\lambda\rightarrow 1)$
in Fig. \ref{fig:ene-D}.
It is seen that two curves drawn for the ground-state energy
intersect each other near $J''/J\sim0.5$,
which implies that a first-order quantum phase transition occurs 
between the ordered phases (I) and (II).  These results agree well
 with those obtained  by the exact
diagonalization, shown by open circles in Fig. \ref{fig:ene-D}
 although there are slight differences between them 
 due to intrinsic deviations for each method.
The critical value for the first-order transition is estimated 
rather accurately, allowing us to determine the phase boundary. 
\begin{figure}[htb]
\begin{center}
\includegraphics[width=7cm]{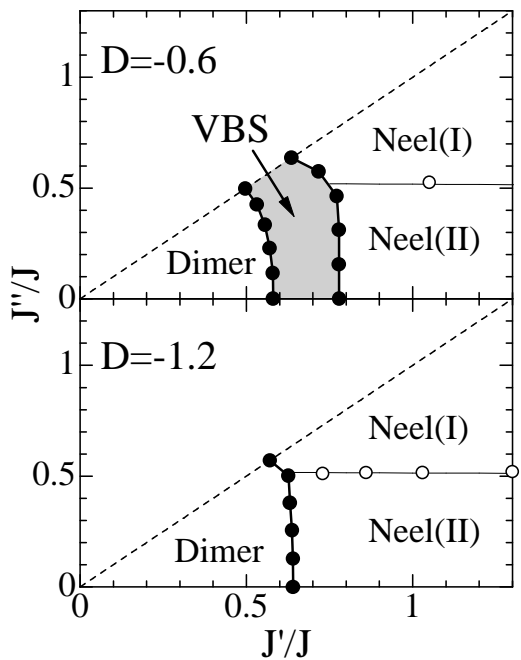}
\end{center}
\caption{
The phase diagram for $D/J=-0.6$ and $-1.2$.
The bold lines with solid circles represent the phase
boundary of the first-order transitions determined by the 
exact diagonalization for the $4 \times 4$ lattice.
The thin solid line between the two Neel phases, which is 
also first-order,  is  determined by the third-order
series expansion of the ground-state energy.  The open circles are
obtained by the exact diagonalization calculation.

}
\label{fig:mix}
\end{figure}

Shown in Fig. \ref{fig:mix} is the phase diagram  thus obtained by the exact
diagonalization and the Ising expansion.
It is shown that the dimer phase and the dimer-plaquette VBS phase
are stable against not only the interchain coupling but also
reasonably large single-ion anisotropy ($D/J=-0.6$).  However, 
when the anisotropy  becomes larger ($D/J=-1.2$), 
the VBS phase eventually disappears in the phase diagram, 
while the dimer phase still exists.
This is consistent with the fact that any VBS-like state may not be
realized  when the system approaches the classical limit.

As for the magnetic states,  there are two 
distinct phases (I) and (II)   separated 
by the first-order phase transition, as mentioned above.
It is to be noted that the
corresponding phase boundary is given by an almost straight line.
This reflects the fact that the chosen anisotropy parameters
$D/J=-0.6$ and $-1.2$ are considered to be rather large in the 
sense that the system possesses the nature expected for a
classical system;
the phase boundary indeed becomes exactly straight for the 
 Ising model. 
The phase diagram shown above for rather large values
of $D$ possesses three or four distinct phases.
We should recall here that the plaquette phase realized for $D=0$ is 
completely suppressed by the presence of the 
magnetic phase (II) in both of the above 
 cases.
When the anisotropy parameter $D$ decreases from these values, the
plaquette phase can overcome the magnetic phase (II) and be
the ground state in the region of small $J''/J$. 
The phase transition between the plaquette phase and the magnetic
phase (II) may be of second order.  Although
we cannot determine such a second-order phase boundary 
correctly by means of the small-size calculation, 
five distinct phases should  be certainly realized for the 
case of small anisotropy.

The calculation presented here  has been restricted to a rather 
small system size or lower-order perturbation. However,  we believe
that the above phase diagram correctly describes  
the ground state properties of the generalized version of 
the  $S=1$ orthogonal-dimer model.

\section{Summary}\label{sec:summary}
We have investigated quantum phase transitions of the $S=1$
Shastry-Sutherland model with single-ion anisotropy. 
By analyzing a generalized 2D model, we have 
shown how the distinct spin gap phases stabilized in the spin chain 
 persist or disappear in the
2D Shastry-Sutherland model. 
The obtained phase diagram has a quite rich structure
especially in the case with single ion anisotropy.
We wish to particularly emphasize
that the frustration-induced VBS phase
exists even in the 2D Shastry-Sutherland model. If this type
of the spin-gap phase can be found in real compounds,
it serves as a novel example of the VBS state in 
2D quantum spin systems. Similar but slightly different 
 VBS singlet phases may be also possible, e.g.
  in the frustrated $S=1$ spin systems 
with the Kagome lattice,\cite{HidaKagome,Uekusa} 
the pyrochlore lattice,\cite{Yamashita,Tsunetsugu,Kogapy} etc.

It is interesting to ask whether a higher-spin Shastry Sutherland
model with $S>1$ can still realize such a VBS state. This is an
open problem to be addressed in the future study.


\section*{Acknowledgement}
We would like to thank H. Kageyama and B. S. Shastry 
for valuable discussions.
This work was partly supported by a Grant-in-Aid from the Ministry 
of Education, Science, Sports and Culture of Japan. 
A part of computations was done at the Supercomputer Center 
at the Institute for Solid State Physics, University of Tokyo
and Yukawa Institute Computer Facility.

\end{document}